# Enhanced Precision in Rainfall Forecasting for Mumbai: Utilizing Physics-Informed ConvLSTM2D Models for Finer Spatial and Temporal Resolution

Ajay Devda[1], Akshay Sunil[2], Murthy R [3]

1. Interdisciplinary Programme in Climate Studies, Indian Institute of Technology, Bombay 2. Department of Civil Engineering, Indian Institute of Technology, Bombay 3. Department of Civil Engineering, BMS College of Engineering

**Abstract**

Predicting precipitation in the tropics has been challenging due to the complex atmospheric dynamics, high humidity, and frequent occurrences of convective rainfall. In the Indian context, the difficulty is further exacerbated because of the monsoon intra-seasonal oscillations which introduce significant variability in rainfall patterns over short periods. Earlier investigations into rainfall prediction leveraged numerical weather prediction methods, along with statistical and deep learning approaches. This study introduces a nuanced approach by deploying a deep-learning spatial model aimed to enhancing rainfall prediction accuracy on a finer scale. In this study, we hypothesise that integrating physical understanding improves the precipitation prediction skill of deep learning models with high precision for finer spatial scales, such as cities. To test this hypothesis, we introduce a physics-informed ConvLSTM2D model to predict 6hr and 12hr ahead precipitation for the coastal city of Mumbai in India. We use the ERA-5 reanalysis data at hourly timesteps from 2011-2022 for the selection of predictor variables, such as temperature, potential vorticity and humidity corresponding to different geopotential levels. The ConvLSTM2D model was trained to the target variable precipitation for 4 different grids representing different spatial grid locations of Mumbai. The NSE metric, utilized to evaluate the precision of 6 and 12 hours ahead precipitation forecasts, yielded ranges of 0.49 to 0.66 for 6-hour predictions and 0.53 to 0.69 for 12-hour predictions during the training phase. In the testing phase, the NSE values ranged from 0.41 to 0.59 for 6-hour forecasts and from 0.41 to 0.61 for 12-hour forecasts, respectively. These values highlight the model's high accuracy and its capacity to capture variations. Thus, the use of the ConvLSTM2D model for rainfall prediction, utilizing physics-informed data from specific grids with limited spatial information, reflects current advancements in meteorological research that emphasize both efficiency and localized precision.

## 1. Introduction

Precipitation, particularly in the form of rainfall, acts as a critical determinant in the initiation and development of natural hazards such as floods, which hugely disrupt human and natural processes. Forecasting precipitation at high resolution hours in advance, supports the real-world socioeconomic needs of many sectors that depend on weather for making critical decisions (Ravuri et al., 2021). Coastal cities are particularly vulnerable to rainfall due to their high economic activities, transportation needs, dense infrastructure and urban sprawl particularly vulnerable to rainfall due to their high economic activities, transportation needs, dense infrastructure, and urban sprawl (Gill and Malamud, 2014; Snow et al., 2012). Rainfall intensity highly varies across small spatial and temporal scales, which creates a necessity to forecast rainfall for finer grids reference to a locality or region within a few kilometres' radius for cities(Liu and Niyogi, 2019; Moron and Robertson, 2020). Short-term, spatially distributed quantitative precipitation forecasting is important to such as flash flood warning systems and operations of dams and reservoirs, and transportation planning(Ali et al., 2014; Yadav and Ganguly, 2020). The temporal and spatial variability of rainfall is highly influenced by a wide range of factors such as atmospheric drivers (atmospheric circulation, etc), temperature, wind patterns, topography, and humidity (Kishtawal et al., 2010; Zhou et al., 2019). These factors control the rainfall behaviour and result in rainfall variation over short distances and time. The chaotic behaviour of atmosphere makes it extremely sensitive to small changes in dynamics leading to vastly different weather phenomenon (Lorenz, 1963). All these complex physical systems along with anthropogenic attributions which cause climate change, make rainfall a complex phenomenon to predict (Allan and Soden, 2008; Trenberth et al., 2003). Considering the dynamics of the Earth system, alongside local atmospheric flows and meteorological traits, aids in accurately depicting the local-scale phenomenon for rainfall predictions (Pielke et al., 1992; Trenberth and Asrar, 2014).

Previous research has investigated relationships between hourly rainfall intensity and different atmospheric variables to understand the choice of predictor variables for precipitation forecast (Lepore et al., 2016; Mitovski and Folkins, 2014; O'Gorman and Schneider, 2009). Lepore et al.

(2016) investigated the relationship between hourly rainfall with humidity, and CAPE (Convective Available Potential Energy) across the United States from 1979 to 2012. This study concluded that moisture availability and vertical instability significantly affect rainfall occurrence. Mitovski and Folkins (2014) analysed high rainfall events across four regions: The Western Tropical Pacific, Tropical Brazil, Southeast China, and the Southeast U.S by utilising 13 years of rainfall data from the Tropical Rainfall Measuring Mission to study high rain events in various regions. This study concluded that mass divergence, relative vorticity, and potential vorticity are related to the dynamics of the atmosphere, affecting convective activity, which can lead to rainfall. Temperature, humidity and vorticity at different geopotential heights affect the cloud formation and considered as potential predictor for rainfall forecasting (Bansod, 2005).

Previous meteorological research has adopted different methods including numerical weather modelling (Holton James R. and Hakim Gregory J., 2012; Warner, 2011), statistical approaches (Fischer et al., 2012; Wilks, 2011) and deep learning techniques (Castro et al., 2021; Espeholt et al., 2022; Shi et al., n.d.; Sønderby et al., 2020; Yadav and Ganguly, 2020) to improve the precipitation forecasts ranging from hourly to annual scales. However, there are a lot of challenges involved in numerical weather prediction models in the tropics due to the inherent complexity and variability of weather patterns which are caused by a myriad of factors, including the El Niño Southern Oscillation (ENSO), the Indian Ocean Dipole (IOD), and the Madden-Julian Oscillation (MJO), each of which can alter the distribution and intensity of monsoonal rains (Ray et al., 2022). This calls for a need to employ modern computing capabilities to enhance the prediction skill of various climate variables (Castro et al., 2021). Recent advancements in computational capabilities such as GPUs and parallel computing, have significantly enhanced the feasibility of employing deep learning techniques in atmospheric and climate sciences research (Lecun et al., 1998; Sønderby et al., 2020; Yadav and Ganguly, 2020). Designing deep neural network architectures for atmospheric systems in specific regions is a complex task, demanding a deep understanding of atmospheric physics and extensive experimentation (Kreuzer et al., 2020; Tong et al., 2022). The complexity is further amplified by the numerous hyperparameters within the network(Dehghani et al., 2023; Espeholt et al., 2022; Sønderby et al., 2020; Xiao et al., 2019), requiring a delicate balance between theoretical knowledge and empirical testing to achieve accurate atmospheric modelling. Prior studies have shown that hybrid models that blend deep learning with physical science enable more precise local rainfall predictions (Castro et al., 2021; Gao et al., 2021). A combination of convolution neural networks (CNN) and long short-term memory (LSTM) has been combined to predict rainfall in spatial and temporal dimensions (Castro et al., 2021; Dehghani et al., 2023; Khan and Maity, 2020; J. Li et al., 2023; W. Li et al., 2022; Yadav and Ganguly, 2020). Castro et al. (2021) utilised a novel deep-learning architecture named STConvS2S to learn the spatiotemporal predictive patterns using meteorological variables such as temperature, pressure, humidity, and wind speed for accurate precipitation forecasting. Suleman and Shridevi (2022) employed a Spatial Feature Attention-Based LSTM (SFA-LSTM) model which has the capability for nuanced spatial and temporal data analysis for short-term weather forecasting. In this study, weather variables include temperature, dew point, relative humidity, station pressure, sea pressure, and wind speed for predicting temperature.

Convolutional LSTM networks integrate the strengths of Convolutional Neural Networks (CNNs) and Recurrent Neural Networks (RNNs) to capture both spatial and temporal information which enhances the accuracy of weather prediction compared to traditional machine learning models(Nastos et al., 2014; Qi et al., 2019; Rasp et al., 2018; Wu et al., 2020; X. Zhang et al., 2018). Gao et al. (2021) reviewed the different deep learning-based methods for precipitation nowcasting and found that ConvLSTM outperforms as compared to other methods. Several attempts have been made to utilize the ConvLSTM to predict rainfall (A. Kumar et al., 2020; B. Kumar et al., 2021; Shi et al., 2017; Yadav and Ganguly, 2020). Yadav and Ganguly (2020) highlight the challenges in predicting very short-term distributed quantitative precipitation forecasting due to atmospheric dynamics. In this study, the authors explore suitable architecture the ConvLSTM model to improve short-term rainfall forecast using NASA's North American Land Data Assimilation System (NLDAS) over the United States. Yasuno et al. (2021) predicted 6-hours ahead rainfall using ConvLSTM for Japan. The study utilized 37 thousand data points hourly radar images from 2006 to 2019 to train and validate the model. The model achieves high RMSE and MAE accuracy which proves the rainfall prediction capability of ConvLSTM for finer spatial scales and therefore ConvLSTM might be more suited to detect sudden changes in the precipitation field if it has seen similar patterns in the training dataset. Multi-source data, in contrast, provide multi-modal and multiscale meteorological information, giving the model a more holistic view of the system. Sønderby et al. (2020) introduced 'MetNet', a neural weather model to forecast precipitation 8 hours ahead using radar and satellite data for the USA. Espeholt et al. (2022) presented 'MentNet-2' a physics-based deep-learning model to predict 12-hour rainfall at 1 km spatial resolution for the USA. The model takes temperature, humidity, wind speed and direction, pressure, and satellite imagery as predictor variables. In the Indian context, Khan and Maity (2020) hybrid Deep Learning (DL) model combining a one-dimensional Convolutional Neural Network (Conv1D) with a Multi-Layer Perceptron (MLP) for predicting daily rainfall up to 5 days in advance. This study used 9 meteorological variables associated with the precipitation as predictors for twelve locations in Maharashtra.

Mumbai is located on the Arabian coast and is frequently exposed to urban floods, socioeconomic loss, and expose to extreme precipitation. Western Ghats, the barriers at western coast, arrest the southwest monsoon rainfall and bring intensified rainfall to the city making it difficult to accurately predict Mumbai's rainfall (Mohanty et al., 2023). Numerous investigations have examined the rainfall patterns and extremity in Mumbai. Gope et al. (2016) presented a Stacked Auto-Encoder (SAE) based deep-learning model to forecast heavy rainfall in Mumbai and Kolkata 6 to 48 hours ahead. The study used surface and at various atmospheric pressure levels (850, 600, and 400-hPa), temperature, mean sea level pressure, relative humidity, and wind velocities (U-wind and V-wind) to predict the rainfall. However, the study restricted extreme events and no spatial variability captured. Previous studies have highlighted the need for exploring and integrating atmospheric variables in deep-learning models to improve precipitation forecasts across different geographical locations and climatic conditions (Castro et al., 2021). Building upon this, in the present study, we hypothesize that that physics-informed deep learning, through the integration of carefully selected atmospheric variables atmospheric variables can improve the rainfall forecast skills. We test this hypothesis for the Coastal City Mumbai, tropical zone where there is high variability in rainfall. Monsoon seasonal oscillations occur as a result of significant seasonal changes in wind patterns, leading to rainfall, particularly during the Southwest monsoon. Mumbai receives a significant portion of its rainfall during this period. Mohanty et al. (2023) identified the three systems responsible for Mumbai's extreme rainfall behaviours, viz, offshore trough, Mid-tropospheric cyclones, and Bay of Bengal (BoB) Depression, which highlights atmospheric variables as the primary influencers of the city's precipitation patterns. The strategic geographic and atmospheric zone with a combination of sea surface and land interactions and monsoon dynamics. There has been no research which has tested the value of physics-informed deep learning methods to forecast rainfall for urban cities with limited spatial atmospheric variables (in Mumbai). Understanding the relationships and the right choice of predictor climate variables has been utilized to train the ConvLSTM2D model which is a state-of-the-art deep learning model to predict rainfall at an hourly scale in this study. The ConvLSTM2D can learn from both spatial and temporal patterns of data and is utilised for learning information from the predictor variables.

The model helps to the significance of physical-based variables influencing the rainfall to deep learning methods. This combination can be further used to provide accurate forecasts of rainfall events at finer spatial scales across Mumbai and other similar settings to help prevent the loss and improve micro-level planning and action by agencies. The remaining of the paper is divided into three sections, methods (section 2), where we provide a detailed overview of the study area, highlighting the geographical and climatic context of Mumbai, followed by model description where we explain the model architecture of the ConvLSTM2D model. Section 3 discuss the model performance in prediction 6h and 12 hr ahead rainfall forecast. Section 4 concludes by summarizing how the ConvLSTM2D model improves rainfall prediction in Mumbai, highlighting its potential for better forecasting in similar vulnerable regions.

## 2. Methods

### 2.1 Study Area

Mumbai is located in the western Ghats region, which belongs to the tropical monsoon climate zone. The city extends between 18.00°– 19.20°N and 72.00°– 73.00°E covering a total area of 437.79 km$^2$. The annual average precipitation is 2450 mm (Mohanty et al., 2023; Singh et al., 2017). Mumbai experienced an unexpected 944mm of rainfall in a single day in the year 2005 (Singh et al., 2017) which cause high lives and economic loss. Such high rainfall extremes in a short period are the main cause of urban floods; a timely and accurate projection of rainfall gives planning agencies enough time to prevent waterlogging and save human and economic losses. Therefore, the crucial necessity of robust nowcasting system arises for coastal cities like Mumbai that can provide the location and intensity of rainfall making people and the public disaster response system well prepared.

Mohanty et al. (2023) highlighted that Mumbai's extreme rainfall events (ERF), with occurrences like the 944 mm downpour on 26 July 2005, are influenced by three key rain-bearing systems: offshore troughs, mid-tropospheric cyclones (MTC), and Bay of Bengal depressions. These systems are instrumental in shaping the city's rainfall patterns, leading to ERF events exceeding 204.5 mm per day approximately once every other year during the summer monsoon season (Gope et al., 2016; Singh et al., 2017; Zope P. E. and Eldho, 2012). This connection underscores the significant impact of these atmospheric phenomena on the frequency and intensity of Mumbai's monsoon extremes. The city is physio-geographically prone to various natural and human-induced hazards, including floods, cyclones, earthquakes, and landslides. Due to high and unplanned urbanization, low laying areas and conjunction in the drainage system, during monsoons almost every year city faces water logging which affects transportation and sometimes leads to loss of lives and economic damage. The rainstorms in this region have the characteristics of high intensity and short duration, as well as long duration for disastrous rainstorms.

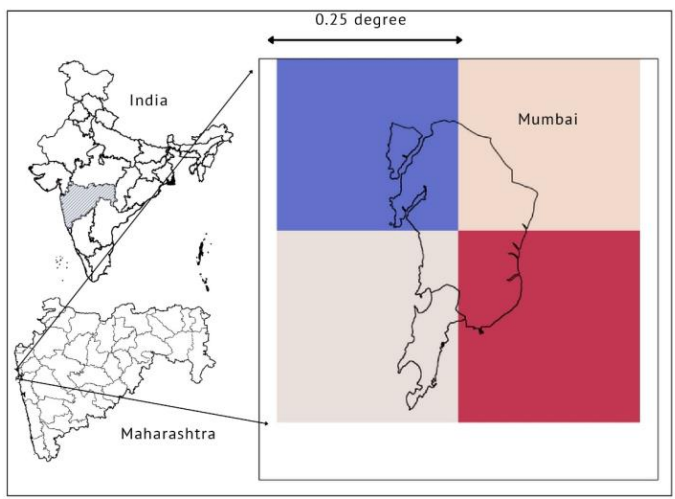

Figure. 1 Location of Mumbai and the grids covered by ERA5 reanalysis products

**2.2 Data**

**2.2.1 Identifying Physics-Informed Variables for High-Resolution Spatial Forecasting of Near-Term Precipitation**

To develop a physics-informed deep learning model for precipitation prediction, it is necessary to understand atmospheric processes such as evaporation, cloud formation, cloud condensation, atmospheric circulation, and the interactions between different layers of the atmosphere to choose appropriate variables as predictors for the deep learning model (Kashinath et al., 2021; Shen and Lawson, 2021; Teufel et al., 2023). The variable selections have been done by a deep literature review of studies related to rainfall predictions and deep learning methods. O'Gorman and Schneider (2009) highlight the importance of considering both specific humidity and temperature changes to accurately predict precipitation extremes, as these variables directly impact cloud condensation processes and the overall atmospheric stability, leading to variations in precipitation intensity and distribution. Weyn et al. (2020) investigated the capability of CNNs in forecasting weather patterns, focusing on the 500-hPa geopotential height as a key input variable. Their research illustrates that CNNs, trained on historical weather data, not only surpass traditional forecast benchmarks but also deliver reliable weather forecasts (500-hPa geopotential height) up to 14 days ahead. Studies have also compared numerical weather prediction (NWP), with deep convolutional neural network (CNN) algorithm finding that CNNs can significantly enhance forecasting accuracy for severe convective weather events (Zhou et al., 2019). In this study, the focus is on developing a deep learning model that combines convolutional layers for spatial analysis and LSTM for temporal analysis to predict rainfall variability and timing (e.g., 6, and 12 hours ahead) at finer spatial scales, such as cities. This model aims to aid meteorological research and assist water resource managers, disaster response teams, and decision-makers in metropolitan areas like Mumbai. It also seeks to provide valuable insights for the urban community and farmers, contributing to a decision support system for rainfall nowcasting. In past studies, many predictors have been used for precipitation downscaling, such as geopotential height (Kidson and Thompson, 1998), sea level pressure (Cavazos, 1999), geostrophic vorticity (Wilby and Wigley, 2000), or wind speed (Murphy, 1999). The choice of the predictors varies across different regions, characteristics of large-scale atmospheric circulation, seasonality, and geomorphology (Anandhi et al., 2009).

2.3 **Data used**

Rainfall is highly sensitive to atmospheric processes such as temperature gradients, vapour and cloud formation and movement (Lenderink and Fowler, 2017). ERA5 reanalysis data is available through ECMWF (European Centre for Medium-Range Weather Forecasts) considering the most appropriate variable, attributing the rainfall gives us a better representation of spatial and temporal diversity at the accuracy level. The atmospheric product of the ERA reanalysis is prepared by complex data assimilation and reconstruction of satellite products and ground observation from weather stations and radars (Hersbach et al., 2020). The dataset provides hourly information on a variety of meteorological variables on regular latitude-longitude grids at 0.25° × 0.25°, such as precipitation, temperature, wind speed, and more, on a global scale covering the period from 1979 to the present (Jiao et al., 2021; Taszarek et al., 2021). Research has underscored enhancements in the ERA5 dataset's spatial and temporal resolution over its predecessor, ERA-Interim, establishing it as a crucial asset for predictive modelling and analytical tasks (Jiao et al., 2021; Nogueira, 2020). Therefore, ERA5 reanalysis datasets have been widely used in climate change research (Letson et al., 2021; Singer et al., 2021; R. Zhang et al., 2021). Previous studies have also highlighted that ERA-5's precipitation data demonstrates notable accuracy and outperforms other reanalysis products at different temporal and spatial scales for India (Mahto and Mishra, 2019).

In this research, we employ data from the ERA 5 reanalysis, selecting 11 predictor variables for forecasting precipitation. Hourly records of these physics-informed variables were downloaded and compiled for the duration spanning from 2011 to 2022, providing a substantial dataset for analysis, model training, prediction generation, and subsequent validation. Table 1 gives comprehensive details of the selected variables in this study. The "total precipitation (tp)" which is the sum of convective and large-scale rainfall (Terblanche et al., 2022). The selected as target variable for the ConLSTM2D model. Previous studies have used ERA-5 "tp" as a measure of rainfall(Jiang et al., 2023).

Table 1: Predictor Variables Derived from ERA5 Reanalysis Data

| Variable | Description | Unit | Reference |
|---|---|---|---|
| t (250 hpa) | Temperature at the geopotential height of 250 hPa | K | |
| t (500 hpa) | Temperature at the geopotential height of 500 hPa | K | (Khan and Maity, 2020) |
| t (850 hpa) | Temperature at the geopotential height of 850 hPa | K | |
| rh (250 hpa) | Relative humidity at the geopotential height of 250 hPa | % | (Khan and Maity, 2020; Salaeh et al., 2022) |
| rh (500 hpa) | Relative humidity at the geopotential height of 500 hPa | % | |
| rh (850 hpa) | Relative humidity at the geopotential height of 850 hPa | % | |
| pv (250 hpa) | Potential vorticity at the geopotential height of 250 hPa | K m² kg⁻¹ s⁻¹ | (Sandhya and Sridharan, 2014; Sun et al., 2023) |
| pv (500 hpa) | Potential vorticity at the geopotential height of 500 hPa | K m² kg⁻¹ s⁻¹ | |
| pv (850 hpa) | Potential vorticity at the geopotential height of 850 hPa | K m² kg⁻¹ s⁻¹ | |
| tcc | Total cloud coverage observed | % | |
| hcc | High-level cloud coverage observed | % | |
| sp | Atmospheric pressure at surface level | Pa | (Khan and Maity, 2020) |

### 2.3.1 Data preparation

In this study, we have selected four different grid locations covering the spatial extent of Mumbai. The global ERA-5 reanalysis dataset was downloaded from the 'ERA-5 hourly data on single levels from 1940 to present' (Hersbach et al., 2020). The data was further cropped using the latitude and longitude extents of Mumbai for the 12 years for the period 2011-2022. This resulted in 105192 hourly samples of selected variables which were then utilized for training and testing the ConvLSTM2D model. The data was further normalised to the [0,1] range to scale the units of different atmospheric variables.

### 2.4 Model Architecture

In this study, we employ the ConvLSTM2D network, an extension of the traditional LSTM, which is designed to handle spatiotemporal data more effectively by incorporating convolution operations in both input-to-state and state-to-state transitions. This integration allows the model to capture spatial dependencies alongside temporal ones, making it particularly suited for tasks like precipitation nowcasting (Espeholt et al., 2022; Gao et al., 2021; Shi et al., 2017; Sønderby et al., 2020). Unlike standard LSTM layers that process data one point at a time, ConvLSTM layers use convolution operations within the LSTM cell, making them suitable for spatial data with temporal dependencies. The model adapted Hierarchical Feature Learning where the first layer captures wide range of features and second layer with fewer filters refine these features and focus on most relevant data.

The ConvLSTM controls the flow of data inside the cell through the forget gate ($f_t$), input gate ($I_t$), and output gate ($O_t$), and the Forget gate decides how information should forget and how information should model retain. (At the end of each iteration, forget gate devices to discard or transmit the relevant information). The memory cell ($C_{t-1}$) acts as an accumulator for state information in an LSTM which accumulates the information. Several self-parametrized controlling gates are employed by the LSTM to access the cell. Activation of the input gate ($I_t$) accumulates the information to the cell. Therefore, the input gate opens the way to integrate the new data in cell and add information to the long-term memory. The forget gate ($F_t$) aids in forgetting information from the past cell status ($C_{t-1}$). Finally, the output gate ($O_t$) propagates the updated information to the next LSTM cell. Here, the values in the output gate ($O_t$) is multiplied with the updated cell information by putting the updated cell state through a tanh activation function to calculate the hidden state ($H_t$).

The mathematical expression of the ConvLSTM layer for each time step t is as follows:

1. Input gate

$$I_t = \sigma(W_{xi} * X_t + W_{hi} * H_{t-1} + W_{ci} \odot C_{t-1} + b_i)$$

2. Forget gate

$$F_t = \sigma(W_{xf} * X_t + W_{hf} * H_{t-1} + W_{cf} \odot C_{t-1} + b_f)$$

3. Cell state

$$Ct_t = F_i C_{t-1} + I_t \odot \tanh(W_{xc} * H_{t-1} * + W_{hc} * H_{t-1} + b_c)$$

4. Output gate

$$O_t = \sigma(W_{zo} * X_t + W_{ho} * H_{t-1} + W_{co} \odot C_t + b_o)$$

5. Hidden state

$$H_t = O_t \odot \tanh(C_t)$$

Where * denotes the convolution operation and ⊙ is elementwise Hadmard product. The σ is the sigmoid activation function and the W represents the state's weighted connections(A. Kumar et al., 2020; Moishin et al., 2021; Shi et al., 2017). σ ensures the gate activations values ranges from 0 to 1, and tanh is a hyperbolic tangent activation function that introduces nonlinearity on the cell update in the operations, and values range from -1 to +1. The term $W_{xi}, W_{xf}, W_{xc}, W_{xo}$ are the weights for the input ($X_t$) to different gates such as input gate, forget gate, cell state, and output gate. $W_{hi}, W_{hf}, W_{hc}, W_{ho}$ are the weights for the hidden state to input gate, forget gate, cell state, and output gate. $b_i, b_f, b_c, b_o$ are the bias terms for the input gate, forget gate, cell state, and output gate. $H_t$ and $C_t$ are the hidden and cell state for the current timestep, and $X_t$ is input at the current timestep. $H_{(t-1)}$ and $C_{(t-1)}$ are the hidden and cell state for previous timestep.

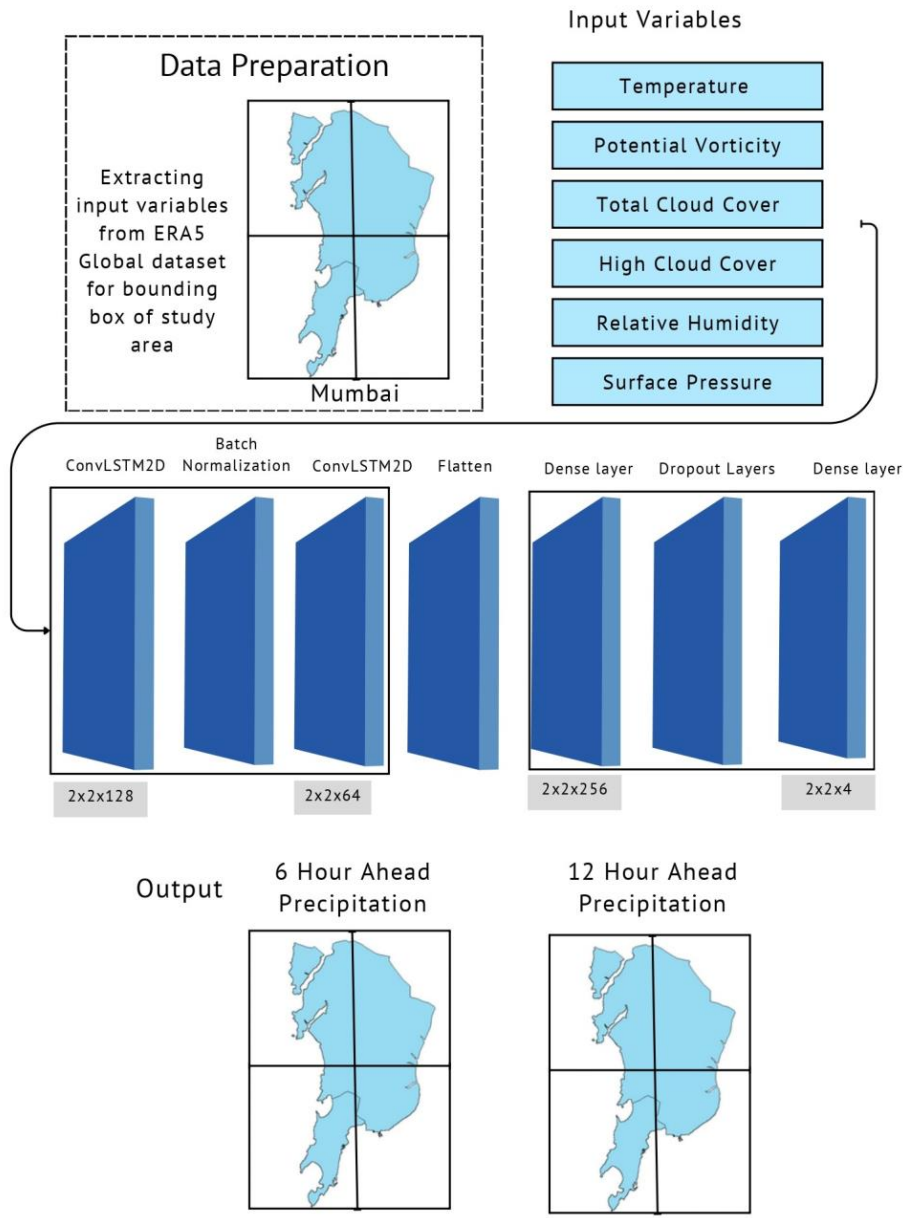

Figure 2. Methodological flow chart for the ConvLSTM2D model

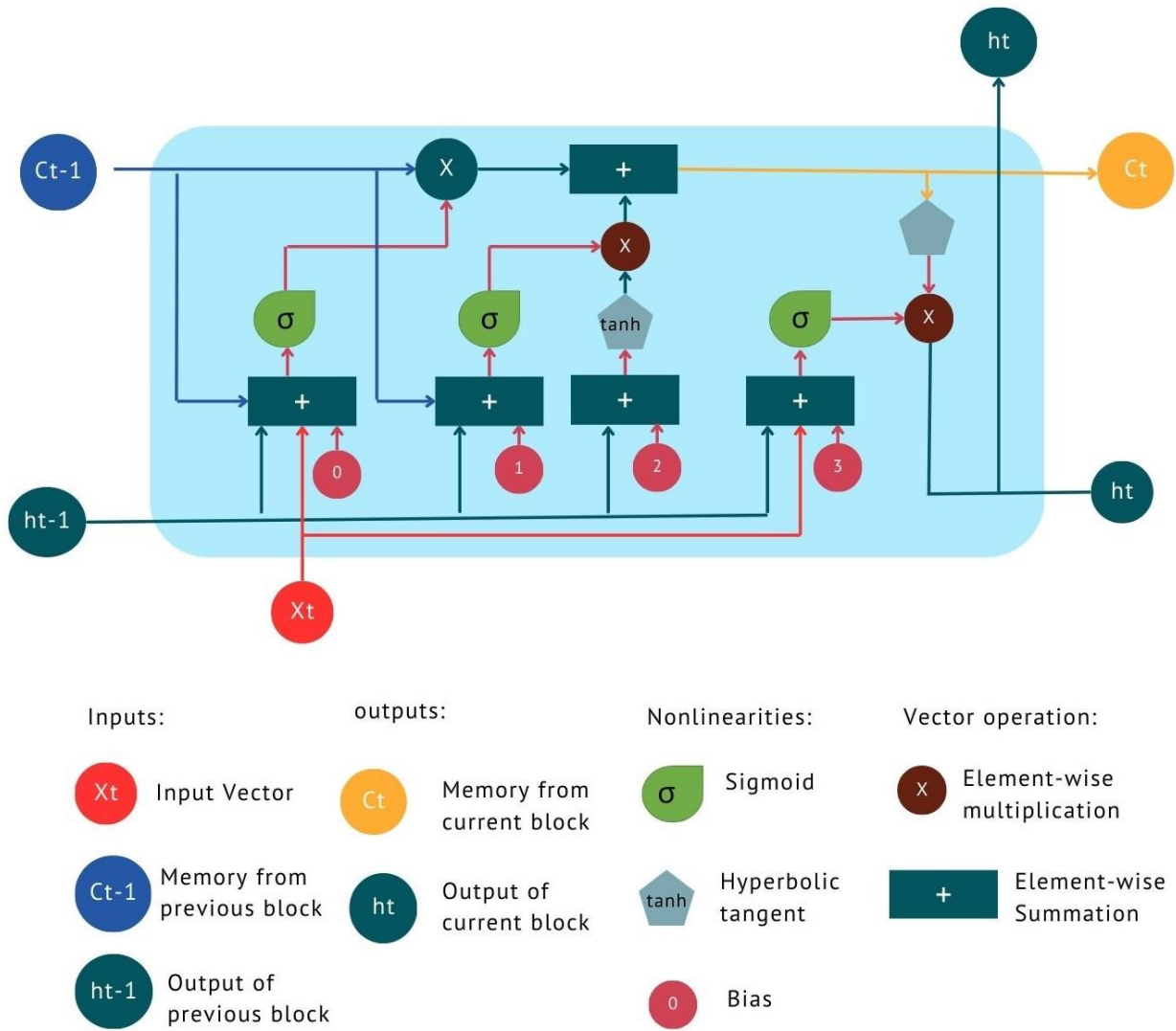

Figure 3. Illustration of the architecture of a single Long Short-Term Memory (LSTM) unit. Each LSTM unit is composed of a cell ($C_t$), an input gate (($I_t$), an output gate ($I_t$), and a forget gate (ft), all of which work together to regulate the flow of information.

The present model used in the study consists of two Conv LSTM layers with different configurations. In a ConvLSTM2D model, the filters extract different features from the input data and captures various aspects from the spatial information (Bi et al., 2023; Guastavino et al., 2022; Ravuri et al., 2021), and therefore a greater number of filters implies model learns from a rich representation of data. The first layer starts the process of extracting relevant features across both dimensions, and the second layer consolidates these features, focusing on the most relevant temporal aspects. In the present model, the first ConvLSTM Layer has with 128 filters and a kernel size of (2, 2). The second ConvLSTM Layer has 64 filters, fewer than the first layer, which helps in progressively reducing the dimensionality of the feature space, making the model computationally efficient. The 'Relu (Rectified Linear Units) ' activation function is employed in the model because of its computational simplicity, introducing non-linearity to the model, allowing it to learn complex patterns. The input shape parameter defines the shape of the input data, which includes the sequence length and spatial dimensions. Finally, the return sequence set to false so the layer will output only to final results to sequins processing, which is suitable for making predictions. The hyperparameters of the ConvLSTM2D model in provided in Table 1.

Table 1. Hyperparameters in ConvLSTM2D architecture

| **Hyperparameters** | **Value** |
|---|---|
| Learning rate | 0.0001 |
| Batch Normalization | True |
| Batch size | 2 |
| Loss function | MSE |
| Activation function | ReLU, Linear |
| Optimizer | Adam |
| Hidden layers | 2 |
| Input data size | 2X2 |
| Number of filters (Input-to-state) | 128 |
| Number of filters (State-to-state) | 64 |
| Kernel size (Input-to-state) | 2X2 |
| Kernel size (State-to-state) | 2X2 |
| Dropout | False |
| Regularizer | False |
| Feature scaling | True [0,1] |

## 2.5 Training and prediction

The current ConvLSTM2D employs a prediction method to generate outputs for both the training and testing datasets, leveraging the learned weights to estimate rainfall patterns from the input features. For the ConvLSTM2D deep learning model, 85% of the dataset, comprising 105,192 hours of predictor and target variables, is allocated to training, with the remaining 15% designated for testing. From the testing dataset, a 15% subset is further designated for model validation. The ConvLSTM2D model processes the dataset initially structured as a 2x2 spatial grid by reshaping it (utilizing reshape(-1, 2, 2)) for compatibility with its training architecture ensuring that the spatial dimensions of the model's outputs align with those of the input grid. Subsequently, the model's performance is evaluated by comparing predictions with the actual data using Root Mean Squared Error (RMSE), Nash-Sutcliffe Efficiency (NSE).

## 3. Result and discussion

In the subsequent sections, we delve into the findings of this study, beginning with an analysis of the correlation matrix for the variables chosen as predictors in the ConvLSTM2D model. We explore the results of both the training and testing phases, assessing the model's performance and its predictive accuracy throughout these stages.

### 3.1 Attribution of variables in rainfall prediction

The correlation and previous literature guide the variable selection for the model, which is discussed in section 2. We selected relevant atmospheric variables and evaluated their correlation to identify predictors for the ConvLSTM2D model, preparing a correlation matrix among these variables to understand their interrelationships and their association with total precipitation (Fig. 4). In the correlation matrix, total precipitation (tp) has a positive correlation with total cloud cover (tcc), high cloud cover (hcc), surface pressure (sp), relative humidity at 850 hPa (rh 850hPa), relative humidity at 500 hPa (rh 500hPa), relative humidity at 250 hPa (rh 250hPa), and potential vorticity at 500hPa (pv 500hPa).

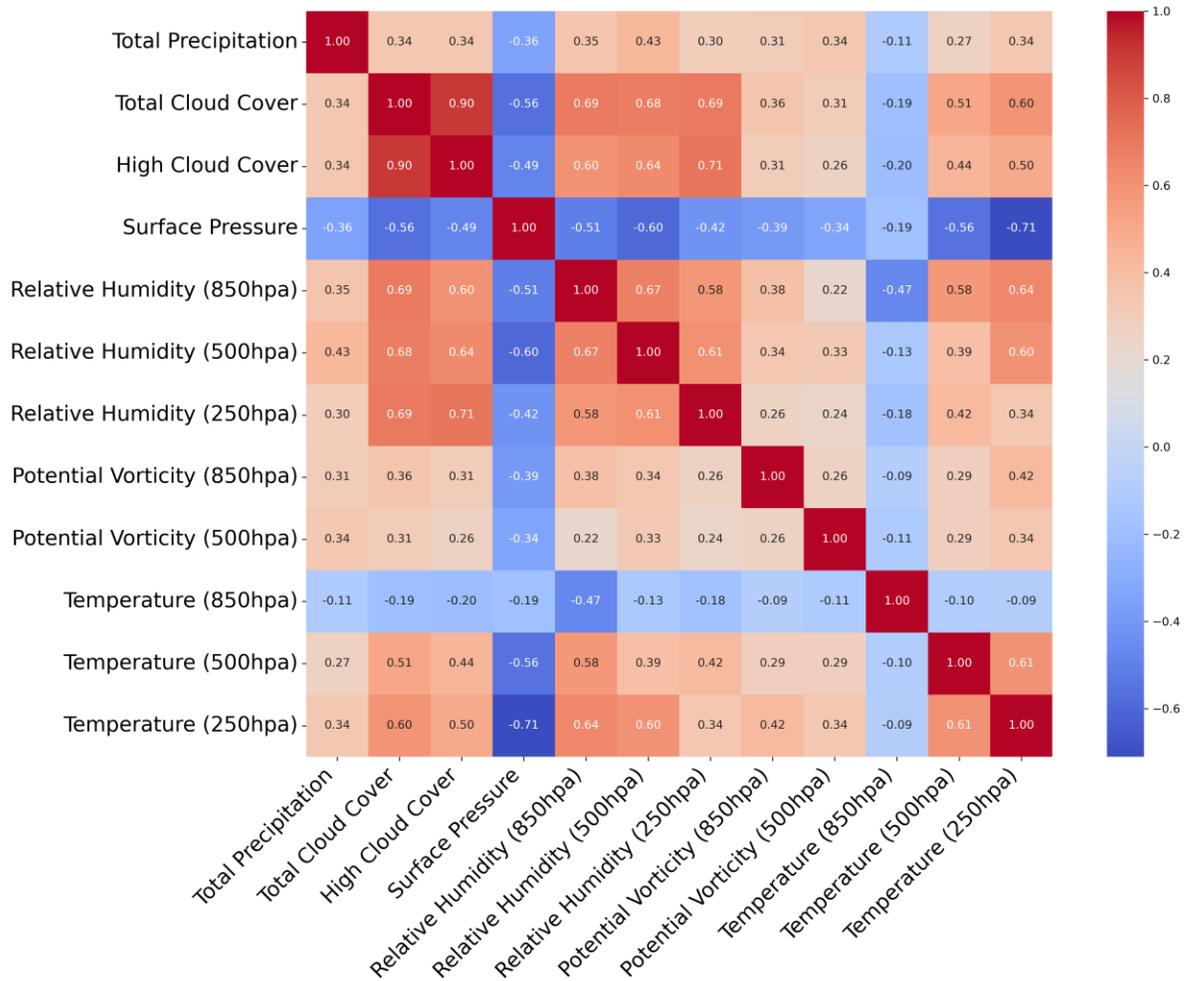

Figure 4. Heatmap representation of correlation matrix of the atmospheric variable where the correlation value varies from -0.71 to 1. The positive correlation values are represented in red while the negative values are represented using the blue colour in the heat map. The correlation for a given variable with itself is 1 as represented using dark red colour in the heat map.

The total cloud cover (tcc) and high cloud cover (hcc) indicate a positive correlation with total precipitation with values of 0.34 and 0.34 respectively which indicates that as the cloud cover increases there is a higher likelihood of precipitation occurrence (Fig. 4). This is consistent with previous studies in atmospheric science which has investigated the relationships between cloud cover and precipitation extremes (Mendoza et al., 2021; Zhong et al., 2021). For example, the presence of high clouds often implies the approach of a front or a low-pressure system, both of which are associated with increased precipitation(Ahrens, 2015). Relative humidity at various tropospheric levels—850hPa, 500hPa, and 250hPa—exhibits positive correlations with total precipitation with correlation values of 0.35, 0.43, and 0.30 respectively, indicating an increasing likelihood of precipitation as relative humidity rises, with the mid-tropospheric level of 500hPa showing the strongest correlation. This is consistent with earlier studies that have emphasized the significant influence of relative humidity on precipitation probabilities. Potential Vorticity at 850hPa and 500hPa shows a positive correlation with total precipitation of 0.31 and 0.34 respectively showing a statistical relationship between higher potential vorticity and temperature. This aligns with previous research which has shown that potential vorticity is an important tracer of air masses as it is conserved in an adiabatic and frictionless flow, making it a useful tracer for air masses(Luhunga and Djolov, 2017; Wallace and Hobbs, 2006) . High potential vorticity is associated with low-pressure systems, leading to enhanced precipitation events(Wallace and Hobbs, 2006). Total precipitation and potential vorticity (PV) are weakly correlated; at geopotential heights of 250hpa. It indicates that while potential vorticity is a measure of the rotation and stratification of the atmosphere, the complex nature of potential vorticity requires conjunction with other factors like humidity and temperature gradients to influence precipitation effectively (Holton James R. and Hakim Gregory J., 2012). Conversely, the correlation between total precipitation and surface pressure is negative ($r = -0.56$), which aligns with established meteorological principles that lower surface pressures are associated with storm systems that can result in increased precipitation events (Davies-Jones and Markowski, 2013). Also, surface pressure influences atmospheric circulation patterns, affecting the spatial distribution and intensity of precipitation. Temperature at different atmospheric levels shows a varied correlation with tp. Interestingly, the temperature at 500 hPa exhibits a weak positive correlation ($r = 0.39$), which may suggest the involvement of warmer air in mid-tropospheric levels contributing to the instability required for cloud formation and subsequent precipitation (White et al., 2016). This could be linked to the concept of convective available potential energy (CAPE), where warmer temperatures aloft can lead to a more unstable atmosphere, potentially resulting in thunderstorms and heavy precipitation under the right conditions (Markowski and Richardson, 2014).

Relative humidity (rh) at different levels shows a moderate to high positive correlation with total precipitation. The correlation of relative humidity at geopotential heights of 250, 500, and 850 hPa is 0.30, 0.43, and 0.34, respectively. This is expected as relative humidity is a direct measure of moisture availability in the atmosphere, which is a critical ingredient for precipitation (Durran and

Klemp, 1982). The hydrolapse, or the decrease in relative humidity with height, may also play a role in determining the likelihood and intensity of precipitation (Emanuel et al., 1994)**.**

**3.2 Comparative Analysis of Predictive Accuracy Across Different Forecast Horizons in Grid-Based Modelling**

In this study, the ConvLSTM2D model has been trained and tested for 6-hour ahead and 12-hour ahead forecasts and performed well for both the cases. In this results section we elaborate on the model performance of 6-hours ahead, and 12-hour ahead forecasts (Fig. 5-8). For the 6-hour ahead forecasts, the results indicate a good correlation between the predicted and actual rainfall values, with correlation (R) ranging from 0.58 to 0.814. The test set correlations are slightly lower, ranging from 0.58 to 0.63. For the training data, the actual maximum rainfall values for the four grids range from 26.74 mm to 32.00 mm, while the model's predictions for these maximum values range from 24.42 mm to 39.40 mm. The actual maximum rainfall values range from 10.27 mm to 17.82 mm for the testing data. The model predictions for these maximum values range from 12.80 mm to 18.54 mm, with the differences between the predicted and actual maximum values ranging from 0.01 mm to 6.18 mm. The actual maximum rainfall values range from 10.27 mm to 17.82 mm for the testing data. The model's predictions for these maximum values range from 12.80 mm to 18.54 mm.

The 12-hour-ahead ConvLSTM2D model demonstrates a commendable performance in forecasting mean rainfall, with its training set predictions spanning from 0.185 mm to 0.271 mm, which is similar to the actual recorded values, which vary from 0.180 mm to 0.269 mm (Fig.7 and 8). The maximum predicted values in the training set, are close to the actual observed maximums, ranging from 26.03 mm to 32.02 mm in predictions versus 26.74 mm to 32.00 mm in actual measurements across the different grids. The correlation between training and testing data sets demonstrates good model performance. The training set correlations between the predicted and actual rainfall values range from 0.68 to 0.77 across the grids. The test set correlations are slightly lower, ranging from 0.68 to 0.73. This shows that the model performed well in all the time ahead forecasts and imitated the actual rainfall with high accuracy. Therefore, our ConvLSTM2D model reveals that the integration of the correct atmospheric variables from a physics-based understanding of the local climate processes with deep learning improves forecast skill in precipitation prediction. The outcomes indicate that the ConvLSTM2D model is well-suited for spatiotemporal forecasting at finer scales, demonstrating commendable accuracy across various lead times.

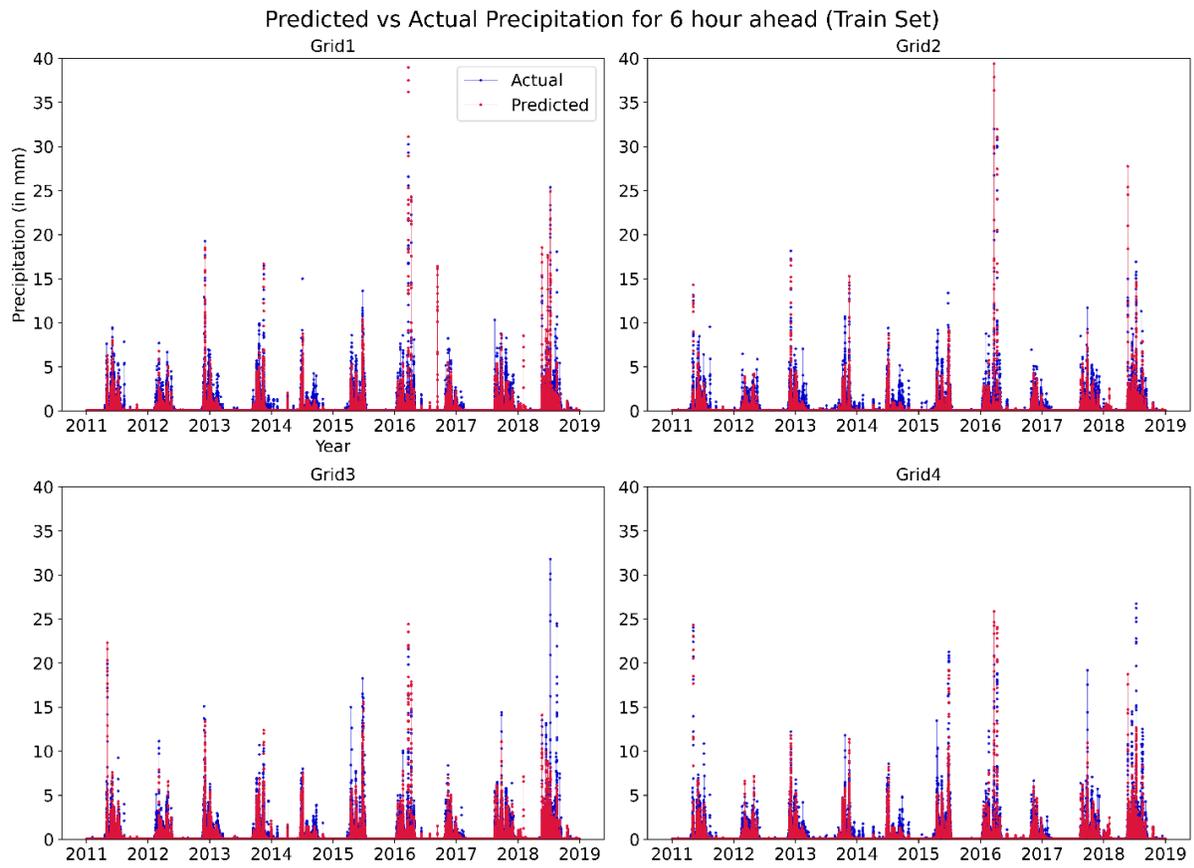

Figure 5. Time series plots representing actual and predicted rainfall for 6 hours ahead for model training. This figure illustrates the time series data from four distinct grid zones, depicting both the forecasted (red dotted line) and the actual (blue dotted line) precipitation values recorded over a multi-year period.

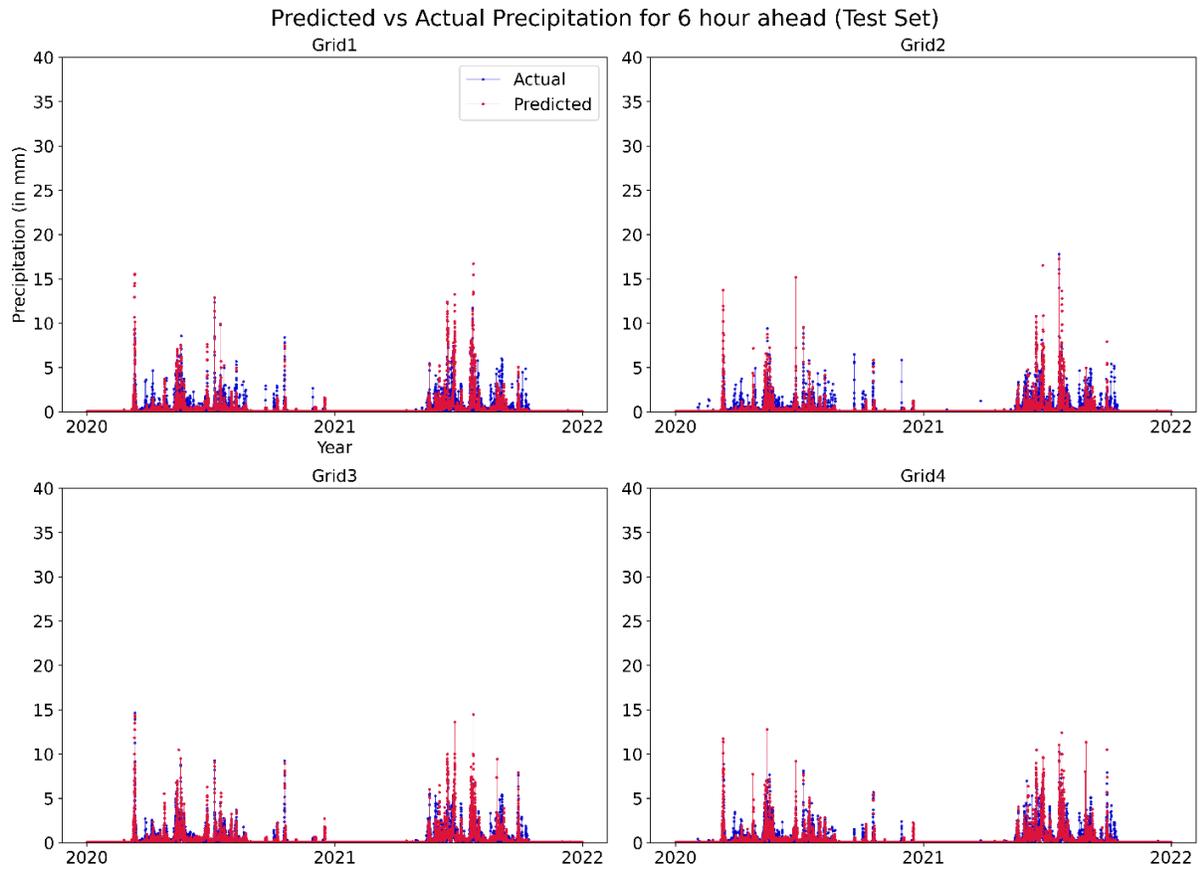

Figure 6. Time series plots representing actual and predicted rainfall for 12 hours ahead for model testing. This figure illustrates the time series data from four distinct grid zones, depicting both the forecasted (red dotted line) and the actual (blue dotted line) precipitation values recorded over a multi-year period.

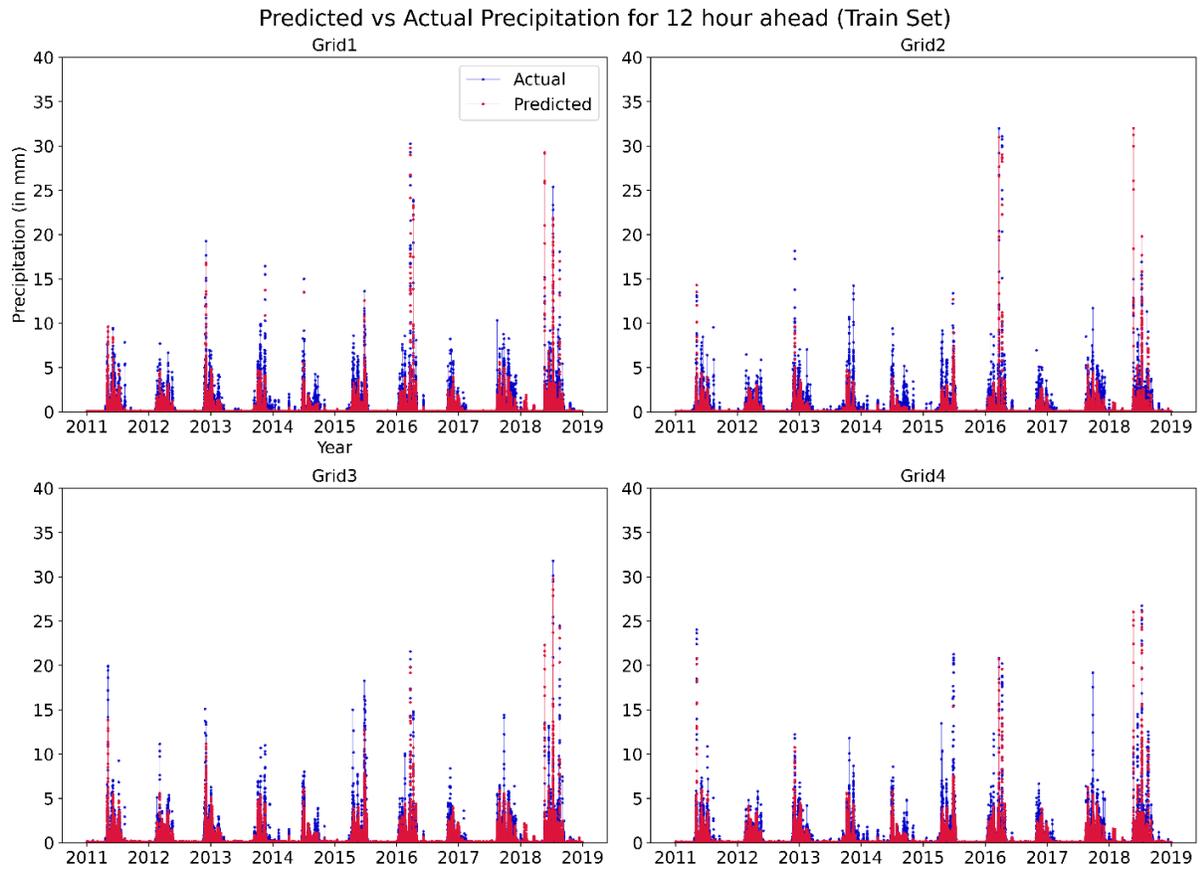

Figure 7. Time series plots representing actual and predicted rainfall for 12 hours ahead for model testing. This figure illustrates the time series data from four distinct grid zones, depicting both the forecasted (red dotted line) and the actual (blue dotted line) precipitation values recorded over a multi-year period.

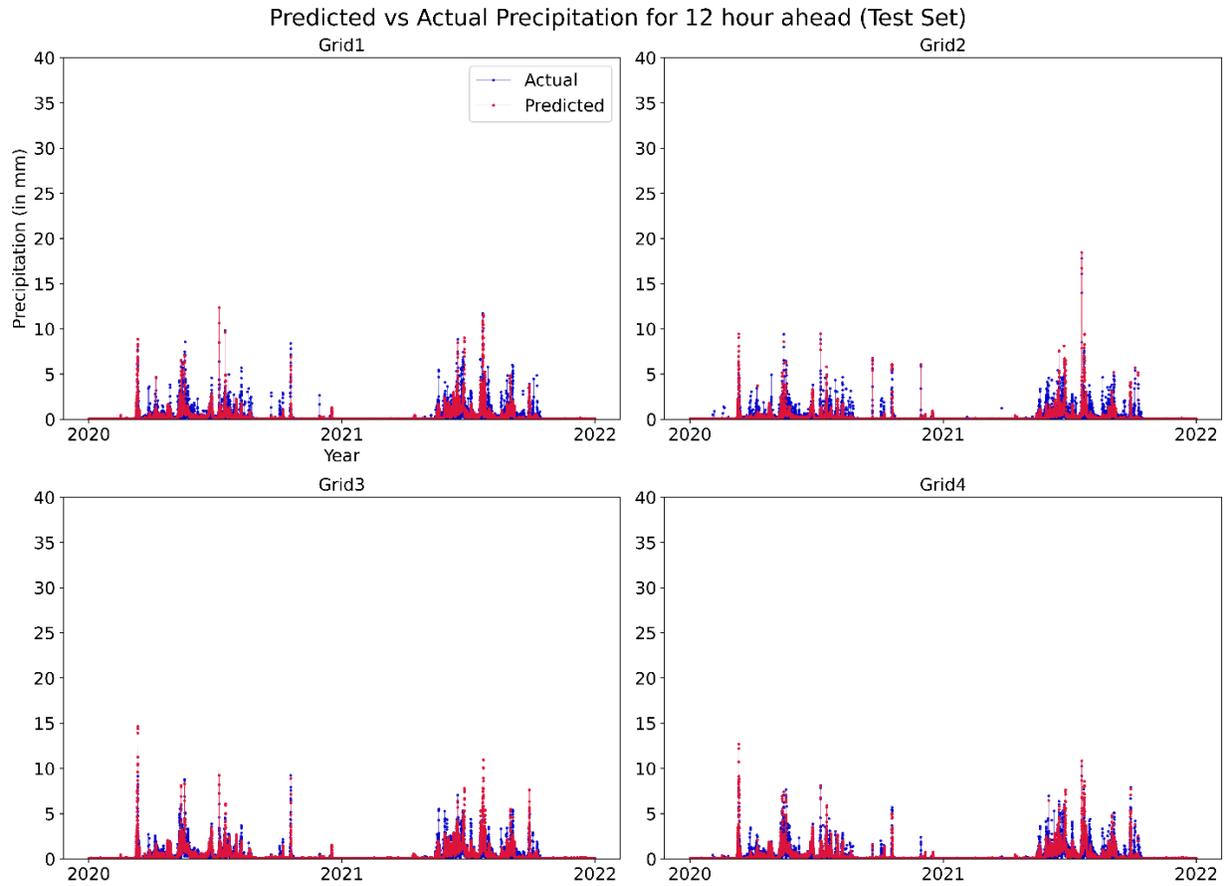

Figure 8. Time series plots representing actual and predicted rainfall for 12 hours ahead for model testing. This figure illustrates the time series data from four distinct grid zones, depicting both the forecasted (red dotted line) and the actual (blue dotted line) precipitation values recorded over a multi-year period.

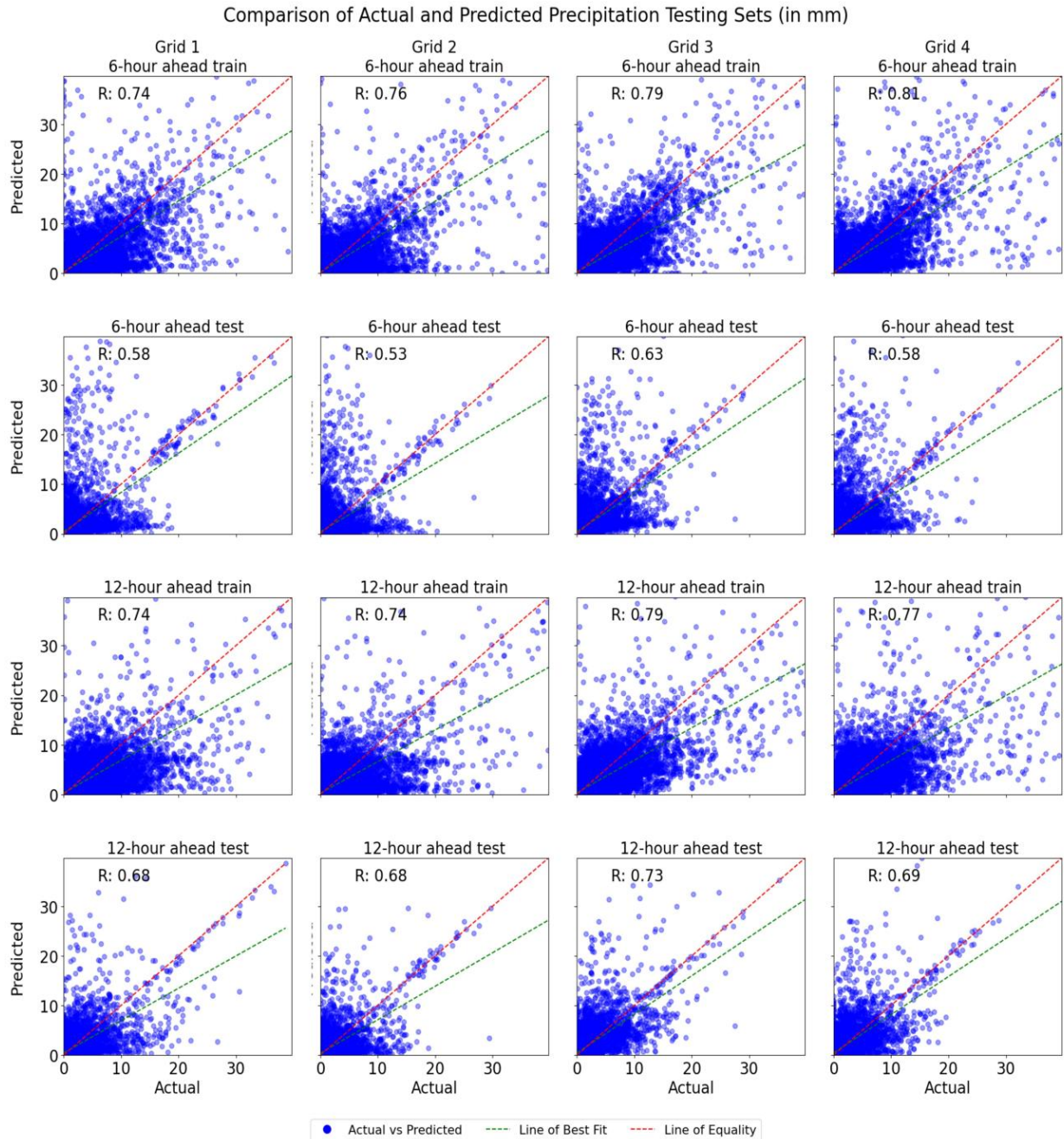

Figure 9: Scatter plots for actual and predicted for (a) 6-hours ahead, and (b) 12-hour ahead for grids1, 2, 3, and 4 on testing the ConvLSTM2D model. The y-axis represents the predicted values, and the x-axis represents the actual values of rainfall.

The correlation plot depicts the relationship between the observed and predicted precipitation over four grids, providing a visual depiction of the model's prediction ability (Fig. 9). Each plot demonstrates the correlation (R) between the observed and forecasted data. The ConvLSTM2D model, averaged across all grids, yields R values of 0.685 for real-time, 0.685 for 2-hour ahead, 0.7 for 6-hour ahead, and 0.54 for 12-hour ahead predictions. The with an overall average $R^2$ of approximately 0.6525 across all grids and time frames.

The correlation plot depicts the predictive performance of the ConvLSTM2D model across four different spatial grids, evaluated at various forecast intervals: 6-hour ahead, and 12-hour ahead predictions. The R values across these grids and timeframes shows a range of model accuracies, with real-time predictions generally yielding the highest R values, indicating a stronger correlation between the predicted and actual values.

**3.2 Evaluation of Model Performance and Nash-Sutcliffe Efficiency (NSE) Metrics**

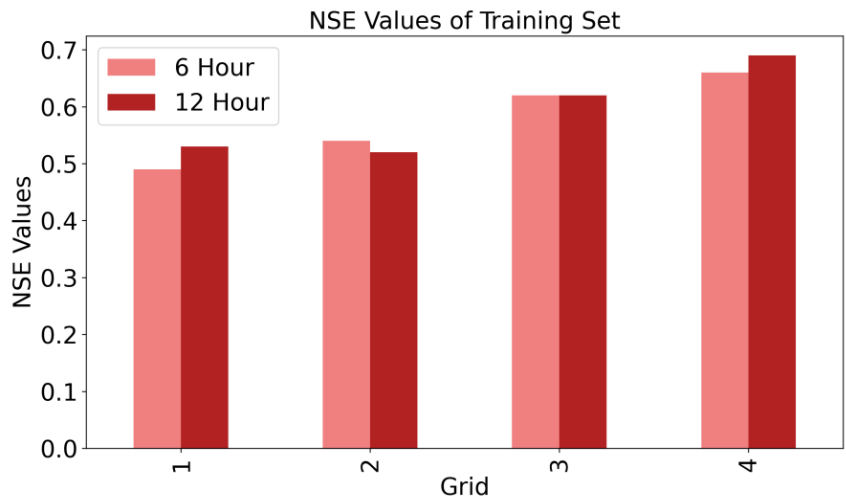

Figure 10. NSE values for training for 6-hour, and 12 hours ahead model training.

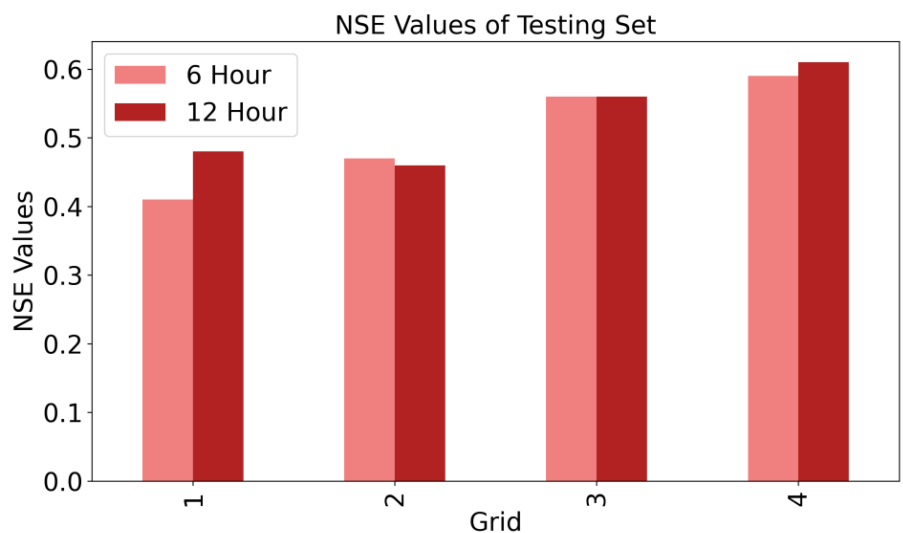

Figure 11. NSE values for training for 6-hour, and 12 hours ahead model testing.

The predictive accuracy is further measured by NSE, across all grids (Fig. 10-11). The NSE values indicate the model is learning with high accuracy and captures the variations. An NSE value of 1 indicates perfect prediction, values closer to 1 indicate superior predictive performance, and a value of 0 corresponds to the model's predictions being no better than the average of the observed data. An NSE value of 1 indicates perfect prediction, suggest better model predictions, 0 indicates the model's accuracy. The NSE values of 6 hours ahead predictions vary from 0.49 to 0.66 for training and 0.41 to 0.59 for testing. Similarly, The NSE values of 12 hours ahead predictions vary from 0.53 to 0.69 for training and 0.41 to 0.61 for testing.

The total precipitation (tp) is the target variable for the ConvLSTM2D model, for predicting future rainfall events based on predictor atmospheric variables. We considered four grids for Mumbai with each grid containing data spanning 105,192 hours, from 00:00 hours on January 1st, 2011, to 23:00 on December 31st, 2022. In this timeframe, 2017, and 2020 experienced flooding in Mumbai. For Grid 1, the highest rainfall recorded was 31.3 mm. This peak rainfall happened on August 29, 2017, at 14:00 hours. In Grid 2, the maximum rainfall recorded was 33.1 mm, observed in the same year as Grid 1, 2017, at 15:00 hours on August 29th. In Grid 3, the peak rainfall of 32.9 mm was recorded on August 5th, 2020, at 09:00 hours. Lastly, Grid 4, recorded its highest rainfall of 27.7 mm on August 5th, 2020, at 09:00 hours, coinciding with the peak rainfall time of Grid 3.

Previous research that relied on numerical weather prediction models encountered limitations in defining broad domains, defining initial conditions and boundary conditions(Hess and Boers, 2022; Schultz et al., 2021). These constraints frequently created complexity and uncertainty in the forecasting process (Warner et al., 1997) . Deep learning models, such as the ConvLSTM2D model discussed in our study, overcome these constraints by effectively utilising data without the need for extensive domain specifications or detailed boundary and initial conditions, bypassing the extensive domain and detailed conditions needed by numerical approaches, thus offering a more direct and potentially more accurate method for localized weather event predictions. One limitation of this study is that the scale of predictions should narrowed down to lower spatial resolutions to make informed decisions at the urban scale. Therefore, the predictors and predictand variables should be available at a finer grid resolution, preferably below

0.25°. Therefore, future research should focus on improving precipitation prediction at finer resolution with available information on the input variables at coarser resolution. This can be accomplished by developing a physics-informed machine learning model that downscales precipitation using the input predictors. This could be a future scope of this research.

**Training and validation loss over epochs**

In the present study runs the model with 200 epochs for all four times of forecast. One epoch represents the one cycle of the data through the ConvLSTM2D layers. This decrement in training loss signifies that the ConvLSTM2D model is effectively learning the spatial-temporal dynamics of rainfall prediction from the training dataset. Validation loss shows slight fluctuations, which is typical in model validation phases due to the model being tested on unseen data. Overall, the decreasing loss functions match the consistency of a well-functioning deep learning model to capture spatial and temporal dynamics of rainfall prediction. A deep learning model with constantly decreasing losses demonstrates constant learning and is considered an effective model.

**Conclusions**

This study is an initial exploration of physics-informed deep learning-based rainfall prediction for fine spatial and temporal scales. The study used ConvLSTM2D, which is suitable for capturing spatial and temporal dimensions to learn and predict rainfall. We developed the appropriate hyperparameters and model structures suitable for predicting rainfall with high accuracy. We found that physics-based variables associated with rainfall occurrence can be a potential predictor as they represent the rainfall phenomena. Moreover, we demonstrate that the input of such variables into the ConvLSTM2D model is capable of predicting rainfall. Deep learning-based models learn from the previous data, and ample data is sufficient to train the model with high prediction skills. However, in the case of the numerical model, initial and boundary conditions pass uncertainties associated with them in prediction. Predicting rainfall in tropical regions is extremely difficult due to the complexity of understanding atmospheric behaviour. In India, these challenges are amplified by the erratic nature of monsoon intra-seasonal oscillations, leading to significant fluctuations in rainfall over short durations. Forecasting extreme weather events, such as heavy precipitation, is essential to prevent human and economic losses. Accurate and timely warnings are powerful tools for capacity building in local disaster management and response agencies. We concluded that understanding the physical processes, atmospheric conditions, and associated variables is helpful in developing a well-performed rainfall prediction model. The model can be further integrated with city-level authorities and as a form of community-based weather forecast services to provide rainfall forecasts.